\documentclass[11pt]{article}
\usepackage[utf8]{inputenc}

\usepackage{fullpage}
\usepackage[margin = 2.5cm]{geometry}
\usepackage[hyperindex,breaklinks]{hyperref}
\usepackage{amsmath, amssymb, amsthm}

\usepackage{graphicx}
\usepackage{color}
\usepackage[dvipsnames]{xcolor}

\usepackage{pgfplots}
\pgfplotsset{compat=1.15}
\usepackage{tikz}

\usepackage{algorithmic}
\usepackage{framed}

\usepackage[round]{natbib}

\usepackage{nicefrac,xfrac}

\newtheorem{theorem}{Theorem}[section]
\newtheorem{claim}[theorem]{Claim}

\newtheorem{lemma}[theorem]{Lemma}
\newtheorem{definition}[theorem]{Definition}

\newcommand{\E}{\mathbf{E}}

\newcommand{\ONE}{\mathbf{1}}

\newcommand{\bbR}{\mathbb{R}}

\newcommand{\calA}{\mathcal{A}}
\newcommand{\calB}{\mathcal{B}}

\newcommand{\calR}{\mathcal{R}}

\newcommand{\calT}{\mathcal{T}}

\DeclareMathOperator{\win}{winner}

\DeclareMathOperator{\cost}{cost}

\newcommand{\sign}{\text{sign}}

\tikzset{My Style/.style={shape=rectangle, rounded corners, align=center, draw, top color=white, bottom color=gray!20}}

\title{Random Cuts are Optimal for Explainable $k$-Medians}
\author{Konstantin Makarychev \and Liren Shan}
\date{Northwestern University}
\begin{document}
\maketitle
\begin{abstract}
We show that the \textsc{RandomCoordinateCut} algorithm gives the optimal competitive ratio for explainable $k$-medians in $\ell_1$. The problem of explainable $k$-medians was introduced by Dasgupta, Frost, Moshkovitz, and Rashtchian in 2020. Several groups of authors independently proposed a simple polynomial-time randomized algorithm for the problem and showed that this algorithm is $O(\log k \log\log k)$ competitive.  We provide a tight analysis of the algorithm and prove that its competitive ratio is upper bounded by $2\ln k+2$. This bound matches the $\Omega(\log k)$ lower bound by Dasgupta et al (2020).
\end{abstract}

\section{Introduction}
In this paper, we provide a tight analysis for the 
\textsc{RandomCoordinateCut} algorithm for  explainable $k$-medians clustering. We show that the competitive ratio of this algorithm is $O(\log k)$. The problem of explainable $k$-medians and $k$-means was introduced by~\cite*{dasgupta2020explainable}. The aim of explainable clustering is to represent data in a way easily understandable by humans. 
\cite{dasgupta2020explainable}
proposed to use threshold decision trees to cluster high dimensional data sets.
A threshold decision tree is a binary space partitioning tree with $k$ leaves. Each internal node of the threshold decision tree splits the data into two groups using a threshold cut $(j,\theta)$: on the one side of the cut, we have points $x$ with $x_j\leq \theta$ and on the other side points $x$ with $x_j > \theta$. Thus, every node of the tree corresponds to a rectangular region of the space. A decision tree with $k$ leaves partitions data set $X$ into $k$ clusters, $P_1,\dots,P_k$. 
\cite{dasgupta2020explainable} suggested that we use
the standard $k$-medians and $k$-means objectives
to measure the cost of the threshold decision tree. For $k$-medians, the cost of a threshold decision tree $\calT$ equals
$$
\cost(X,\calT) = \sum_{i=1}^k \sum_{x \in P_i} \|x-\hat c^i\|_1,
$$ 
where $P_1,\dots, P_k$ is the partitioning of $X$ produced by $\calT$; and $\hat c^1,\dots, \hat c^k$ are the medians of clusters $P_1,\dots, P_k$. We denote the $\ell_1$-norm by $\|\cdot\|_1$. Note that each $P_i$ is a rectangular region of the space. Thus, generally speaking, every $x$ is not assigned to the closest center $c_1,\dots,c_k$ like in unconstrained $k$-medians or $k$-means. 

\begin{figure}\label{fig:diagram-example}
    \centering
    \includegraphics[width=0.35\linewidth]{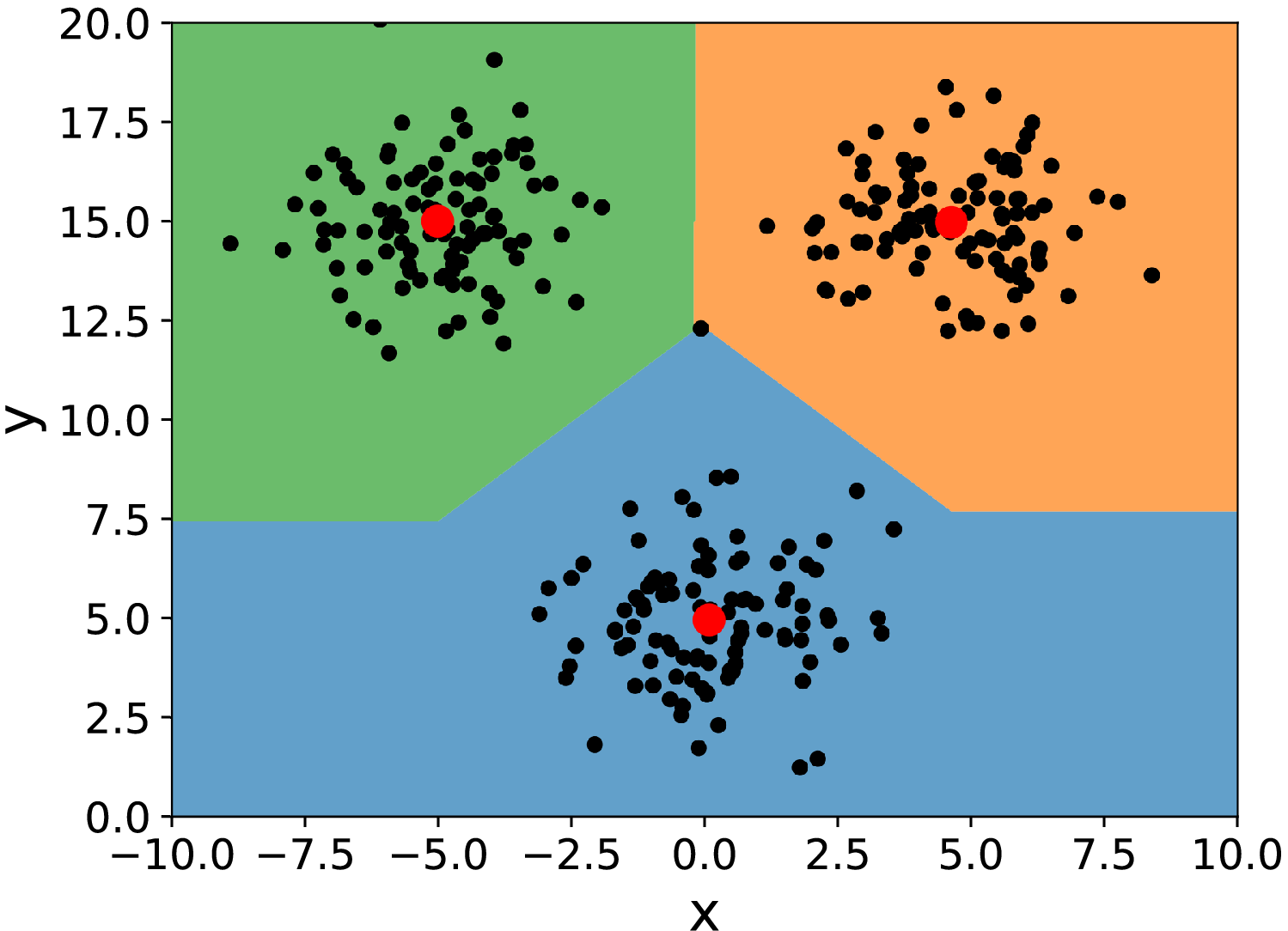}
    \includegraphics[width=0.35\linewidth]{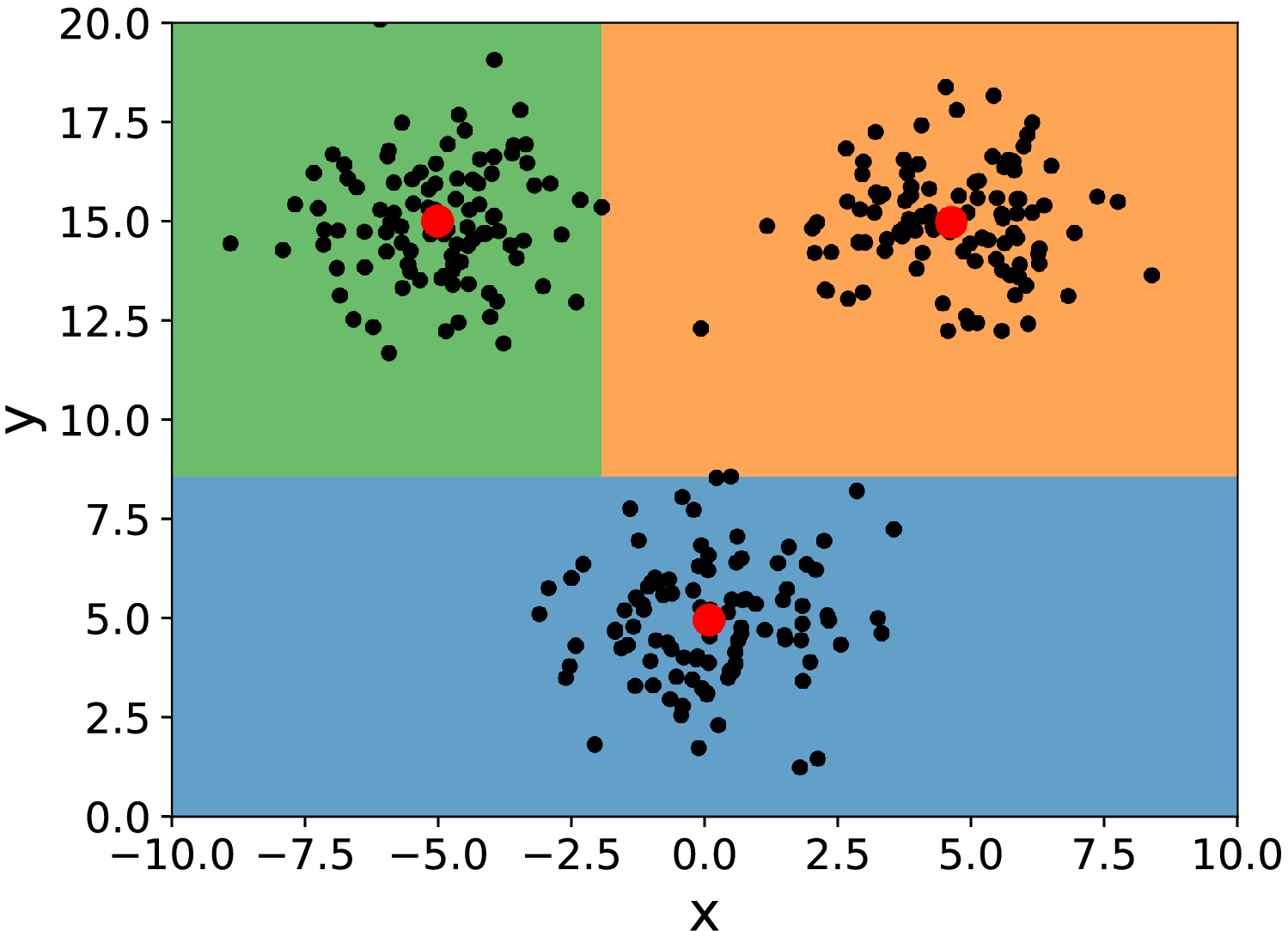}
    \begin{tikzpicture}
    \node[My Style]{$y\leq 8.6$}
    child{node[style={shape=rectangle, rounded corners, align=center, draw, fill=blue!60}]{$1$}}
    child{node[My Style]{$x \leq -1.9$}
    child{node[style={shape=rectangle, rounded corners, align=center, draw, fill=green!60}]{$2$}}
    child{node[style={shape=rectangle, rounded corners, align=center, draw, fill=orange!60}]{$3$}}}
    ;
    \end{tikzpicture}
    \caption{The unconstrained $k$-medians clustering and explainable $k$-medians clustering. The left diagram shows the Voronoi partition of the plane w.r.t. three centers in $\ell_1$ distance. The Voronoi cell for each center consists of all points that are closer (in $\ell_1$ distance) to this center than to any other center  (the boundaries between cells are not straight lines because we use the $\ell_1$ distance). The middle diagram shows an explainable partition. The right diagram shows the corresponding decision tree for explainable clustering.
    }\label{fig:example}
\end{figure}

\cite*{dasgupta2020explainable} defined the price of explainability as the ratio of the $k$-medians  cost of explainable clustering to the optimal cost of unconstrained $k$-medians clustering.
They
showed that the cost of explainability for $k$-means and $k$-medians (somewhat surprisingly) does not depend on the number of points in the data set $X$ and only depends on $k$. Specifically, they provided a greedy algorithm that given $k$ reference centers $c^1,c^2,\cdots, c^k$ of any unconstrained $k$-medians as input, outputs a threshold decision tree of cost at most $O(k)$ times the cost of original unconstrained $k$-medians with centers $c^1,c^2,\cdots, c^k$. We call such an algorithm $O(k)$ competitive. To get an explainable $k$-medians clustering, we first obtain reference centers $c^1,c^2,\cdots, c^k$ using an off-the-shelf 
approximation algorithm for $k$-medians 
and then run an $\alpha$-competitive algorithm for explainable $k$-medians with centers $c^1,c^2,\cdots, c^k$ given as input.
This algorithm produces the desired threshold decision tree.
\cite{dasgupta2020explainable}  also gave an $O(k^2)$ competitive algorithm for $k$-means and
showed  $\Omega(\log k)$ lower bounds on the price of explainability for both $k$-medians and $k$-means. 

The notion of explainable clustering immediately got a lot of attention in the field (\cite{laber2021price,makarychev2021near, gamlath2021nearly, charikar2022near, esfandiari2022almost}). Particularly, \cite*{makarychev2021near, esfandiari2022almost} provided almost optimal algorithms for explainable $k$-medians, and 
\cite*{makarychev2021near, esfandiari2022almost, gamlath2021nearly} provided almost optimal algorithms for $k$-means. 
The competitive ratios of these algorithms are $\tilde O(\log k)$ for $k$-medians and $\tilde O(k)$ for $k$-means. 

The algorithms for explainable $k$-medians by \cite*{makarychev2021near, esfandiari2022almost, gamlath2021nearly} are variants of the same simple algorithm, which we call \textsc{RandomCoordinateCut}. This algorithm receives a set of $k$ reference centers $c^1,\dots,c^k$ as input and then builds a threshold decision tree with $k$ leaves. It works as follows. It recursively partitions $d$-dimensional space until every cell contains exactly one reference center $c^i$. The algorithm starts with a tree consisting of one node, the root. Initially, all $k$ reference centers are assigned to that root. At every step, the algorithm picks a random threshold cut $(j,\theta)$ and splits centers in every cell using this cut. 
If this cut does not separate any centers in a cell $u$ (i.e., all centers in $u$ are located on one side of the cut), then the algorithm does not split $u$ into two regions at this step. Finally, for every leaf $u$ of the constructed tree, the unique center that belongs to the cell corresponding to $u$ is assigned to $u$. 
We provide pseudo-code for this algorithm in Figure~\ref{alg:threshold_tree_kmedians}.

\cite{makarychev2021near, esfandiari2022almost} showed that the competitive ratio of \textsc{RandomCoordinateCut} is at most $O(\log k\log\log k)$. That is, for every data set $X$ and set of centers $c_1,\dots,c_k$,
$$
\E[\cost(X,\calT)] \leq O(\log k\log\log k) \cdot \sum_{x\in X} \min_{c\in\{c_1,\dots,c_k\}}\|x-c\|_1.
$$ 
Note that the running time of this algorithm is $\tilde{O}(kd)$. 
\cite*{gamlath2021nearly} provided a slightly worse bound of $O(\log^2 k)$ on the competitive ratio  of this algorithm. They also conjectured that this algorithm is optimal and its competitive ratio is $O(\log k)$, more specifically, $H_{k-1}+1$, where $H_k$ is the $k$-th harmonic number. They provided some justification for their conjecture by proving this bound for a very special set of centers and data points (corresponding to the case of completely disjoint sets in our Set Elimination Game).

\begin{figure}[tb]
\begin{framed}
\begin{algorithmic}
\STATE {\bfseries Input:} a data set $X \subset \bbR^d$ and set of centers 
$C=\{c^1,c^2,\dots, c^k\} \subset \bbR^d$
\STATE {\bfseries Output:} a threshold tree $\calT$
\STATE {}
\STATE Create tree $\calT_0$ containing a root node $r$. Assign $C_r = \{c^1,c^2,\cdots,c^k\}$ to the root.  Let $t=0$. 
\STATE Let $M = \max_{ij} |c^i_j|$. 
\STATE {}
\WHILE{$\calT_n$ contains a leaf with at least two distinct centers}
\STATE Pick a random coordinate $j$ and random $\theta \in (-M,M)$. Let 
$\omega_n = (j,\theta)$.
\STATE For every leaf node $u$ in $\calT_n$, split the set $C_u$ into two sets: 
$$
\text{\emph{Left}} = \{c \in C_u: c_j \leq \theta\}
\text{\;\;\;\;\;\;\;\;\;\; and\;\;\;\;\;\;\;\;\;\;}
\text{\emph{Right}} = \{c\in C_u: c_j > \theta\}.$$ 
If both sets are not empty, then create two children of $u$ in tree $\calT_t$. The left child corresponds to the subregion of $u$ with $x_j\leq \theta$, and the right child corresponds to the subregion of $u$ with $x_j > \theta$.
Assign sets \emph{Left} and \emph{Right} to the left and right child, respectively.
\STATE {}
\STATE Denote the updated tree by $\calT_{t+1}$.
\STATE Update $t = t+1$.
\ENDWHILE
\end{algorithmic}
\end{framed}
\caption{\textsc{RandomCoordinateCut} algorithm}
   \label{alg:threshold_tree_kmedians}   
\end{figure}

\textbf{Our Results.} In this work, we show that indeed the competitive ratio of \textsc{RandomCoordinateCut} is at most $2\ln k+2$, and, therefore, this algorithm has the optimal competitive ratio which matches the lower bound of \cite*{dasgupta2020explainable}. Our analysis is not only tight but also fairly simple. To get our result we define a game, the Set Elimination Game, which was also implicitly analyzed in previous works on this topic. We show that the cost of this game is at most $2\ln k+2$.

\textbf{Related Work. }
The unconstrained $k$-medians clustering has been extensively studied. 
\cite*{charikar1999constant} gave the first constant factor approximation algorithm for the problem in general metric spaces.
\cite{li2013approximating} provided a $1+\sqrt{3}+\varepsilon$ approximation algorithm. \cite*{byrka2017improved} improved the approximation factor to $2.675+\varepsilon$. \cite*{cohen2022improved} recently improved the approximation factor to $2.406$ for Euclidean $k$-medians.
\cite*{megiddo1984complexity} showed that the $k$-medians in $\ell_1$ problem is NP-hard.
\cite*{cohen2022johnson} showed that it is also NP-hard to approximate $k$-medians in $\ell_1$ within a factor of $1.06$. 

As we discuss above, 
\cite*{gamlath2021nearly},
\cite*{esfandiari2022almost}, 
\cite*{makarychev2021near}, independently proposed the
\textsc{RandomCoordinateCut} algorithm. They also gave an $ \tilde O(k)$ algorithm for explainable $k$-means and showed a lower bound of $\tilde \Omega(k)$ for the problem. 
\cite{charikar2022near} provided an $O(k^{1-2/d}\cdot \mathrm{poly}(d,\log k))$ competitive algorithm for explainable $k$-means, whose competitive ratio  depends on the dimension $d$ of the instance. For small $d \ll \log k / \log\log k$, their bound is better than $O(k)$. They showed an almost matching  $\Omega(k^{1-2/d}/\mathrm{ploy}\log k)$ lower bound for explainable $k$-means. 
\cite{esfandiari2022almost} gave an upper bound of $O(d\log^2 d)$ on the competitive ratio of  \textsc{RandomCoordinateCut} for explainable $k$-medians. This bound is better than $O(\log k)$ for small $d \ll \log k / \log\log k$.
\cite{laber2021price} gave $O(d\log k)$ and $O(dk\log k)$ competitive algorithms for explainable $k$-medians and $k$-means, respectively. 
\cite*{frost2020exkmc} provided some empirical evidence that bi-criteria algorithms for explainable $k$-means (that partition the data set into $(1+\delta)k$ clusters) can give a much better competitive ratio than $O(k)$. Then,
\cite*{makarychev2022explainable} gave a $\tilde O(\frac{1}{\delta}\log^2 k)$ competitive bi-criteria algorithm for explainable $k$-means. \cite*{bandyapadhyay2022find} provided an algorithm that computes the optimal explainable $k$-medians and $k$-means clustering in time $n^{2d+O(1)}$ and $(4nd)^{k+O(1)}$, respectively. \cite*{laber2023shallow} proposed to use  shallow decision trees for explainable clustering.
\section{Set Elimination Game}\label{sec:set-elimination-game}

In this section, we define the set elimination game. Consider a finite measure space $(\Omega,\mu)$ and $k$ distinct sets $S_1,S_2,\dots, S_k \subset \Omega$. These sets $S_1,S_2,\dots,S_k$ may overlap with each other. The set elimination game proceeds in a series of rounds. Initially, all sets $S_1,\dots,S_k$ enter the competition. Formally, they belong to the set of remaining sets $\calR_0 = \{S_1,\dots,S_k\}$. At every round $n$, the host picks a random $\omega_n\in \Omega$ with probability $\Pr(\omega_n =\omega)= \mu(\omega)/\mu(\Omega)$. Then, all sets $S_i$ that contain $\omega_n$ are eliminated from the game unless all remaining sets contain $\omega_n$, in which case, no set gets eliminated. That is, for $n\geq 1$,
\begin{equation}\label{eq:def:remaining-sets}
\calR_{n} = 
\begin{cases}
\calR_{n-1}\setminus
\{S_i\in \calR_{n-1}: \omega_n \in S_i\},&\text{if for some }S_i\in \calR_{n-1}, \omega_n\notin S_i;  \\
\calR_{n-1},&\text{otherwise}.
\end{cases}
\end{equation}
The last remaining set is declared the winner. We denote that winner by $\win$. We say that the cost of the game is the measure of the winning set, $\mu(\win)$.

We remark that $\calR_n$ cannot get empty (in which case, the winner would not be defined) because of the ``otherwise'' clause in the definition~(\ref{eq:def:remaining-sets}). We shall always assume that all sets $S_1,\dots, S_k$ are not only distinct and non-empty but also (a) for every $i$, $\mu(S_i)>0$, and (b) for all $i$ and $j$, $\mu(S_i \triangle S_j) > 0$ (here, $S_i \triangle S_j$ denotes the symmetric difference of sets $S_i$ and $S_j$). Then, in every game, there is a unique winner with probability $1$. 

Our main result is the following theorem, which, as we discuss later in Section~\ref{sec:reduction}, implies that the competitive ratio of the explainable clustering algorithm is $2\ln k+2$.

\begin{theorem}\label{thm:main}
Consider a set elimination game with the finite measure space $(\Omega, \mu)$ and $k$ distinct sets $S_1,S_2,\dots, S_k$ (as above). The expected cost of the game is at most
$$
\E\big[\mu(\win)\big] \leq 
(2\ln k +2) \cdot \min_{i \in [k]} \mu(S_i).
$$
\end{theorem}

To simplify the exposition, we will prove this theorem for discrete finite measure sets. If $\Omega$ is not a discrete measure space, we first replace it with a quotient space: We say that $\omega'\in \Omega$ and $\omega''\in \Omega$ are equivalent ($\omega'\sim \omega''$) if they are contained in exactly the same set of sets $S_1,\dots, S_k$. This equivalence relation partitions $\Omega$ into at most $2^k$ different equivalence classes. We  replace $\Omega$ with the quotient space $\nicefrac{\Omega}{\sim}$ whose elements are equivalence classes. In other words, we merge all equivalent $\omega$'s. The measure of a new element $\tilde\omega$ equals to the measure of the corresponding equivalence class.

\textbf{Organization.}
In
Section~\ref{sec:reduction}, we discuss the connection between explainable $k$-medians and set elimination games. We define a set elimination game in a set system $I\subset \{S_1,\dots,S_k\}$ in Section~\ref{sec:local-compete}. Then, we define the hitting and elimination time in Section~\ref{sec:exp-clock}. We illustrate our proof strategy by showing Theorem~\ref{thm:main} for the case when the smallest set $S_1$ does not overlap with $S_2,\dots,S_k$ in Section~\ref{sec:disjoint}. An important ingredient of our proof is the notion of \emph{surprise sets}, which we discuss in Section~\ref{sec:surprise}. Finally, we complete the proof of Theorem~\ref{thm:main} in Section~\ref{sec:general-case}.

\subsection{Explainable $k$-Medians via Set Elimination Game}\label{sec:reduction}

In this section, we show how to use  Theorem~\ref{thm:main} to obtain a bound of 
$2\ln k + 2$ on the competitive ratio of the \textsc{RandomCoordinateCut} algorithm.

\begin{theorem}\label{thm:explain_clustering}
The competitive ratio of the \textsc{RandomCoordinateCut} algorithm
for Explainable $k$-Medians is at most
$2\ln k +2$. That is, for every set of centers $C=\{c_1,\dots,c_k\}$ and data set $X$, the algorithm finds a random decision tree $\calT$ such that
$$
\E[\cost(X,\calT)] \leq (2\ln k+2) \cdot
\sum_{x\in X}
\min_{c\in\{c_1,\dots,c_k\}}\|x-c\|_1.
$$
\end{theorem}
The pseudo-code for the \textsc{RandomCoordinateCut} algorithm is provided in Figure~\ref{alg:threshold_tree_kmedians}.

\begin{proof}
Consider an arbitrary data set $X \subset \bbR^d$ and set of $k$ centers $C \subset \bbR^d$. We assume that all points in $X$ and all centers in $C$ are in the cube $[-M,M]^d$.  
The threshold decision tree obtained by the  \textsc{RandomCoordinateCut} algorithm partitions the space into $k$ cells. Each cell contains a single reference cluster $c^i$. The center $c_i$ is not necessarily optimal for cluster $P_i$ (cluster $P_i$ is the intersection of the data set $X$ and $i$-th cell). However, we will use it as a proxy for the optimal center. In other words, we will upper bound the cost of the threshold decision tree as follows:
$$
\cost(X,\calT) \equiv \min_{\hat c^1,\dots,\hat c^k}\sum_{i=1}^k \sum_{x \in P_i} \|x-\hat c^i\|_1
\leq
\sum_{i=1}^k \sum_{x \in P_i} \|x-c^i\|_1.
$$

Let $\Omega$ be the set of all coordinate cuts:
$\Omega = \{(j,\theta): j \in [d], \theta \in [-M,M]\}$. We define a measure $\mu$ on  $\Omega$ as follows. For every subset $S \subset \Omega$, we set
$$
\mu(S) = \sum_{i=1}^d \mu_L(\{\theta: (j,\theta) \in S\}),
$$
where $\mu_L$ is the Lebesgue measure on $\bbR$. 
Thus, we have $\mu(\Omega) = 2dM$, which implies $(\Omega,\mu)$ is a finite measure space.

Consider any data point $x \in X$. Define $k$ sets $S_1,S_2,\dots,S_k$ for the set elimination game. For every $i \in \{1,\dots,k\}$, let $S_i$ be the set of all threshold cuts that separate $x$ and center $c^i$, i.e.,
$$
S_i = \{(j,\theta)\in \Omega : \sign(x_j - \theta) \neq \sign(c^i_j - \theta) \}.$$
Note that the $\ell_1$ distance from $x$ to center $c^i$ equals  the measure of $S_i$: $\|x-c^i\|_1 = \mu(S_i)$.  We now examine the set elimination game with sets $S_1,\dots, S_k$, measure space $(\Omega,\mu)$, and random sequence of draws $\omega_1,\omega_2,\dots$ (each $\omega_n\in \Omega$ is the threshold cut chosen by the \textsc{RandomCoordinateCut} algorithm at step $n$). We claim that $S_i$ belongs to $\calR_n$ if and only if center $c_i$ lies in the same cell as point $x$ after step $n$ of the algorithm.
This is the case for $n=0$, since $\calR_0$ contains all sets $S_1,\dots, S_k$ and the root of the threshold tree contains all centers $c^1,\dots,c^k$. Then, whenever we pick cut $\omega_n$, all centers separated from $x$ by $\omega_n$ are removed from the cell of $x$. The only exception from this rule occurs when all centers in that cell lie on the same side of the cut $\omega_n$. That is exactly the same rule as we have for the set elimination game (note that center $c^i$ is separated from $x$ by $\omega_n$ if and only if $\omega_n \in S_i$). Therefore, the same sets $S_i$ remain in the game as center $c^i$ in the cell of $x$ (namely, sets $S_i$ and centers $c^i$ have the same indices).

The \textsc{RandomCoordinateCut} algorithm stops when all leaves of the decision tree contain exactly one center. At this step, the set elimination game contains one set, $S_i$. This set corresponds to the center $c^i$ assigned to point $x$.
The cost of the game $\mu(S_i)$ equals  the distance from $x$ to $c^i$. 
By Theorem~\ref{thm:main}, we have
$$
\E[\cost(x,\calT)] 
=\E[\mu(\win)]
\leq (2\ln k +2) 
\cdot
\min_{i}
\mu(S_i)
=
(2\ln k +2)
\cdot
\min_{i}
\|x-c^i\|_1.
$$
We sum this bound over all data points $x$ in $X$ and get the desired result.
\end{proof}

\subsection{Local Competitions}\label{sec:local-compete}
We now revisit the definition of the set elimination game and define competitions in subsets of 
$\{S_1,\dots,S_k\}$. We remind the reader that every set elimination game is determined by an infinite sequence of i.i.d. random variables $\omega_1,\omega_2,\dots$. For each round $n$ and element $\omega\in \Omega$, $\Pr\big(\omega_n = \omega\big) = \mu(\omega)/\mu(\Omega)$. 

\begin{definition} Consider a finite measure space $(\Omega,\mu)$. Let $I$ be a set of subsets of $\Omega$. We say that $I$ is a valid set system if (a) for every $S\in I$, $\mu(S) > 0$, and (b) for every $S',S''\in I$, $\mu(S'\triangle S'') > 0$.
\end{definition}

The reader may assume that $(\Omega,\mu)$ is a discrete finite measure space and $\mu(\omega) > 0$ for all $\omega$ in $\Omega$. Then, the definition above says that in a valid set system $I$, all sets are non-empty and disjoint.

\begin{definition}
Consider a finite measure space $(\Omega,\mu)$. Let $\omega_1,\omega_2,\dots$ be i.i.d. random variables as described above and $I$ be a valid set system. We define a set elimination game in $I$. Initially, $\calR_{0}(I) = I$. Then, for every $n\geq 1$,
\begin{equation}\label{eq:def:remaining-sets-local}
\calR_{n}(I) = 
\begin{cases}
\calR_{n-1}(I)\setminus
\{S\in \calR_{n-1}(I): 
\omega_{n} \in S\},&\text{if for some }S'\in \calR_{n-1}(I), \omega_{n}\notin S';  \\
\calR_{n-1}(I),&\text{otherwise}.
\end{cases}
\end{equation}
The winner of the game in $I$, denoted by $\win(I)$, is the 
only element remaining, or, formally, the unique element in $\cap_{n\geq 0} \calR_n(I)$. If $\cap_{n\geq 0} \calR_n(I)$ contains more than one element, then the winner is not defined. The cost of the game is the measure of the winner,$\mu(\win(I))$.
\end{definition}
We remark that $\cap_{n\geq 0} \calR_n(I)$ contains exactly one element with probability $1$. Thus, the winner and cost of the game are defined with probability $1$.

Consider sets $S_1,\dots, S_k$ from Theorem~\ref{thm:main}. Denote $K=\{S_1,\dots, S_k\}$. The definition of the competition among sets $S_1,\dots, S_k$ (given in the beginning of Section~\ref{sec:set-elimination-game}) is exactly the same as the definition of competition in $K$. Our goal is to show that $\E[\mu(\win(K))] \leq 2(\ln k + 2) \cdot \min_{S_i \in K}\mu(S_i)$.
In the proof of Theorem~\ref{thm:main}, we will consider competitions in different set systems $I\subseteq K$. We prove the following key lemma.

\begin{lemma}\label{lem:winner-in-set-winners}
Consider a partitioning of the set system $K= \{S_1,\dots,S_k\}$ into $m$ sets $I_1,\dots,I_m$. Then,
$$
\win(K) \in 
\big\{\win(I_1),\dots,\win(I_m)\big\}.
$$
\end{lemma}
The proof of Lemma~\ref{lem:winner-in-set-winners} relies on the following observarion.
\begin{lemma}\label{lem:A-t-X-Y}
Let $X$ and $Y$ be two subsets of $K$. If $X\subset Y$, then for every $n$, we always have
\begin{equation}\label{eq:AtY-subset-AtX}
\calR_n(Y)\cap X =
\calR_n(X) \;\;\;\text{ or }\;\;\; \calR_n(Y)\cap X = \varnothing.
\end{equation}
\end{lemma}
\begin{proof}
We prove that~(\ref{eq:AtY-subset-AtX}) holds by induction on $n$. Initially, when $n=0$, we have $\calR_0(X) = X$ and $\calR_0(Y) = Y$. Therefore, $\calR_0(Y)\cap X = X\cap Y = X = \calR_0(X)$. Suppose (\ref{eq:AtY-subset-AtX}) holds for $n$, we prove that (\ref{eq:AtY-subset-AtX}) also holds for $n' = n+1$.
If $\calR_n(Y)\cap X = \varnothing$, then $\calR_n(Y)\cap X$ remains empty for all $n'\geq n$. Therefore,
(\ref{eq:AtY-subset-AtX}) holds for $n + 1$. So, let us assume that $\calR_n(Y)\cap X = \calR_n(X)$. Consider three cases:
\begin{itemize}
\item If $\omega_{n+1}$ belongs to all sets in $\calR_n(Y)$, then it also belongs to all sets in $\calR_n(X) = \calR_n(Y)\cap X$. Thus, in this case, no set is eliminated in $X$ or $Y$. That is, 
$\calR_{n+ 1}(X) = \calR_n(X)$ and 
$\calR_{n+ 1}(Y) = \calR_n(Y)$. 
\item If $\omega_{n+1}$ belongs to all sets in $\calR_n(X)$, but not all sets in 
$\calR_n(Y)$, then, at step $n+1$, we remove all sets that contain $\omega_{n+1}$ and, 
particularly, all sets in $\calR_n(X)$,
from $\calR_n(Y)$. Consequently, $\calR_{n+1}(Y) \cap X = \varnothing$ .
\item If not all sets in $\calR_n(X)$ and not all sets in $\calR_n(Y)$ contain $\omega_{n+1}$, then we remove exactly the same sets from both $\calR_n(X)$ and $\calR_n(Y)\cap X$. Namely, we remove sets $S_i\in\calR_n(Y)$ that contain $\omega_{n+1}$.
\end{itemize}
We conclude that (\ref{eq:AtY-subset-AtX}) holds for $n'=n+1$.
\end{proof}
\begin{proof}[Proof of Lemma~\ref{lem:winner-in-set-winners}]
Consider an arbitrary realization of the game $\omega_1,\omega_2,\dots$. Let $n$ be the round when all sets but the winner are eliminated from the competition i.e., $\calR_n$ contains only one set, the winner. Since $K$ is the union of $I_1,\dots,I_k$, the winner must belong to some $I_j$. Now, by Lemma~\ref{lem:A-t-X-Y} for $X=I_j$ and $Y=K$, we have 
$\calR_n(K)\cap I_j = \calR_n(I_j)$ or $\calR_n(K)\cap I_j = \varnothing$. We know that $\calR_n(K)=\{\win(K)\}$ and $\win(K) \in I_j$. Thus, $\calR_n(K)\cap I_j = \{\win(K)\}\neq \varnothing$, and 
$$\calR_n(I_j) = 
\calR_n(K)\cap I_j = \{\win(K)\}.
$$
We conclude that at round $n$, $\calR_n(I_j)$ contains only one set -- the winner in $K$. Consequently, it is also the winner in $I_j$ i.e., $\win(I_j) = \win(K)$. This finishes the proof.
\end{proof}

\subsection{Set Elimination with Exponential Clock}\label{sec:exp-clock}

Consider a set elimination game on sets $S_1,\dots,S_k$. It is determined by the sequence of random i.i.d. draws $\omega_1,\omega_2,\dots$. Random variable $\omega_n$ is chosen in round $n$. We assign every round a random time $\tau_n$. Let the time between two consecutive rounds be an exponential random variable with parameter $\mu(\Omega)$. Specifically, let $\Delta\tau_1,\Delta\tau_2,\dots$be a sequence of i.i.d. exponential random variables with parameter $\mu(\Omega)$ and each $\tau_n = \tau_{n-1} + \Delta \tau_n = \Delta \tau_1 + \dots+\Delta\tau_n$. Note that all $\Delta \tau_n$ are positive and $\tau_1,\tau_2,\dots$ is an increasing sequence with probability $1$.
The number of draws that occurs by time $t$ (i.e., $N_t(\Omega) = |\{n: \tau_n \leq t\}|$) is a Poisson process with parameter $\mu(\Omega)$. We now can think of the set elimination game as follows: The host of the game observes a Poisson process with parameter $\mu(\Omega)$. Whenever the process jumps (at time $\tau_n$), the host picks an element $\omega_n$ in $\Omega$ with probability $\Pr(\omega_n=\omega) = \mu(\omega)/\mu(\Omega)$ and eliminates some sets according to the rules of the game discussed above. Note that by assigning every round some time $\tau_n$, we do not change the game, the winner, and the cost of the game (because the sequence of random draws $\omega_1,\omega_2,\dots$ remains the same as before). This interpretation of the game allows us to introduce a hitting time $h(S)$ of every subset $S\subset \Omega$ with the following properties: (a) each $h(S)$ is an exponential random variable with rate $\mu(S)$; (b) hitting times of disjoint sets are mutually independent random variables. 

\begin{definition}
For every subset $X \subset \Omega$, the hitting time $h(X)$ is the time 
$\tau_n$ when the first $\omega_n$ is drawn from $X$:
$$
h(X) = \min \{\tau_n: \omega_n \in X\}.
$$
When the set contains one element $\omega$, we will write $h(\omega)$ instead of $h(\{\omega\})$.
\end{definition}
We also define the elimination time of each set $S_i$. 

\begin{definition}
Consider any set elimination game with the measure space $(\Omega,\mu)$ and $k$ sets $S_1,S_2,\dots,S_k$ in $\Omega$. The elimination time $e(S_i)$ of set $S_i$ is the time when set $S_i$ is eliminated from the game, i.e., 
$$e(S_i) = \min \{\tau_n: S_i \notin \calR_n(K)\}.$$ 
If $S_i$ is the winner, then we let $e(S_i) = \infty$ (because the winner is never eliminated).  \end{definition}
Note that $e(S_i)\geq h(S_i)$. Sometimes, $e(S_i)$ may be equal to $h(S_i)$, but 
$e(S_i)$ and $h(S_i)$ are not always the same.
We now prove that hitting times for  disjoint sets are independent. To this end, we \emph{split} the Poisson process $N_t(\Omega) = |\{n: \tau_n \leq t\}|$. Let 
$$N_t(\omega) = |\{n: \tau_n \leq t\text{ and }\omega_n = \omega\}|.$$
It is easy to see that $N_t(\Omega) = \sum_{\omega\in \Omega} N_t(\omega)$ for every $t$. It is also true that each $N_t(\omega)$ is a Poisson process with parameter $\mu(\omega)$ and all $N_t(\omega)$ (for $\omega \in \Omega$) are mutually independent. This fact follows from the Coloring Theorem (see e.g.,~\cite{kingman-poisson}, Coloring Theorem, page 53).
\begin{theorem}[Coloring Theorem]
Let $\Pi_t$ be a Poisson process on the real line with rate $\lambda$. We color each event of the Poisson process randomly with one of $M$ colors: The probability that a point receives the $i$-th color is $p_i$. The colors of different points are independent. Let $\Pi_t(i)$ be the number of events of color $i$ in the interval $(0,t]$. Then, $\Pi_t(1),\dots,\Pi_t(M)$ are independent Poisson processes. The rate of process $\Pi_t(i)$ is $\lambda p_i$.
\end{theorem}

\begin{lemma}\label{lem:exp-time-hitting-time}
For every $\omega\in \Omega$, $h(\omega)$ is an exponential random variable with parameter $\mu(\omega)$, and all random variables $h(\omega)$ (for $\omega\in \Omega$) are mutually independent.
\end{lemma}
\begin{proof}
Observe that $h(\omega) = \min\{t: N_t(\omega)\geq 1\}$. Thus, $h(\omega)$ is an exponential random variable (the time of the first jump of a Poisson process) with rate $\mu(\omega)$. Also, since all $N_t(\omega)$ (for $\omega\in \Omega$) are mutually independent, all $h(\omega)$ are also mutually independent.
\end{proof}

Note that the set elimination game depends only on the hitting times for elements $\omega$ in $\Omega$. This is the case because it matters only when every $\omega$ is drawn the first time. At that time -- the hitting time of $\omega$ -- 
all sets that contain $\omega$ are eliminated unless all remaining sets contain this $\omega$. When the same $\omega$ is drawn again, it does not eliminate any new sets. Also, note that for any set $S\subset \Omega$, the hitting time $h(S) = \min_{\omega\in S}h(\omega)$. Thus, $h(S)$ is an exponential random variable with parameter $\mu(S) = \sum_{\omega\in S}\mu(\omega)$.

\section{Proof of Main Result}
We now present the proof of our main result, Theorem~\ref{thm:main}. We assume without loss of generality that $S_1$ is the smallest set i.e., $\mu(S_1)\leq \mu(S_i)$ for all $i$. Then, the expected cost of the game is at most:
\begin{equation}\label{eq:exp-cost-game-proof}
\mu(S_1) + \sum_{i=2}^k \Pr\big(S_i = \win(K)\big)\mu(S_i).
\end{equation}

\subsection{Special Case: $S_1$ is Disjoint from $S_i$}\label{sec:disjoint}
We first provide some intuition for the proof by considering the case when $S_1$ does not intersect with sets $S_2,\dots,S_k$, i.e. sets $S_1$ and $S_i$ are disjoint for all $i =2,3,\dots,k$. We split all sets into two groups $S_1$ and the rest of the sets $S_2,\dots,S_k$. We know from Lemma~\ref{lem:winner-in-set-winners} that the winner among all sets $S_1,\dots,S_k$ is either $S_1$ or $\win\big(\{S_2,\dots,S_k\}\big)$. 
Denote $I^-=\{S_2,\dots,S_k\}$. Each set $S_i$ is eliminated at time $e(S_i)$. The set $S_1$ is eliminated 
at its hitting time 
$h(S_1)$ unless it is the only remaining set at  time $h(S_1)$ (because we are considering the case when $S_1$ does not overlap with other sets). Thus,
\begin{equation}\label{eq:win-K-warmup-2}
\win(K) = 
\begin{cases}
S_1,&\text{if } h(S_1) > e(\win(I^-));\\
\win(I^-),&\text{if } e(\win(I^-)) > h(S_1).
\end{cases}
\end{equation}
This means that a set $S_i$ is the winner if $S_i = \win(I^-)$ and $e(S_i) > h(S_1)$.
Thus, the cost of the game is bounded by 
$$\mu(S_1) + \sum_{i=2}^k \Pr\Big(S_i = \win(I^-)
\text{ and } h(S_1) < e(S_i) \Big)\cdot \mu(S_i).
$$
We want to argue that $S_i$ cannot compete against $S_1$ if $h(S_1)$ is sufficiently large. Since $S_1$ and $S_i$ are disjoint, when the first set among $S_1$ and $S_i$ is hit, it gets eliminated. Hence, $h(S_1)<e(S_i)$ implies $h(S_1)<h(S_i)$. We know that $\Pr\big(h(S_1)<h(S_i)\big) = \mu(S_1)/\mu(S_1\cup S_i)$ because $h(S_1)$ and $h(S_i)$ are independent exponential random variables. So, if events $\{S_i = \win(I^-)\}$ and $\{h(S_1) < h(S_i)\}$ were independent, we would replace 
$\Pr\big(S_i = \win(I^-)
\text{ and } h(S_1) < e(S_1)\big)$ with the product
$\Pr\big(S_i = \win(I^-)\big)\cdot 
\Pr\big(h(S_1) < h(S_i)\big)$ and easily obtain the desired result. However, these events are not independent. In fact, if $S_i$ is the winner in the set system $I^-$, then its hitting time $h(S_i)$ is likely to be large. Nevertheless, we argue that 
$\min(h(S_1),e(S_i))$ is not likely to be too large (see below for a formal statement). We will need the following definitions.
\begin{definition}\label{def:surprise}
We say that $S_i$ is a surprise set if $e(S_i)\geq h(S_1)\geq L/\mu(S_i)$, where $L=\ln k$. 
\end{definition}

Let us examine bound~(\ref{eq:exp-cost-game-proof}). 
Let $Surprise$ be the set of all surprise sets. Note that $Surprise$ is a random set. Then,
\begin{align}\label{eq:warmup-main}
\sum_{i=2}^k \Pr\big(S_i = \win(K)\big)\mu(S_i)
&\leq \sum_{i=2}^k \Pr\big(S_i = \win(K),\;S_i \notin Surprise\big)\cdot\mu(S_i)\\ \notag
&+\sum_{i=2}^k \Pr\big(S_i \in Surprise\big)\cdot \mu(S_i).
\end{align}
We show in the next section (Lemma~\ref{lem:pr-surprise-set}) that the second sum is upper bounded by $\mu(S_1)$. We now bound the first sum. 
For every winner $S_i$ which is not a surprise set, we have $e(S_i)\geq h(S_1)$ (because $S_i$ is the winner) and $h(S_1)\leq L/\mu(S_i)$ (because $S_i$ is not a surprise set). We also have $S_i= \win(I^-)$, thus
$$\Pr\big(S_i = \win(K),\;S_i\notin Surprise\big)
\leq \Pr\big(h(S_1)\leq L/\mu(S_i)\text{ and } S_i = \win(I^-)\big). 
$$
By Lemma~\ref{lem:exp-time-hitting-time}, all hitting times $h(S_i) = \min_{\omega\in S_i} h(\omega)$ for $i \geq 2$ are independent from  $h(S_1)$. Thus,
$\win(I^-)$ is also independent of $h(S_1)$ ($\win(I^-)$ depends only on the hitting times for sets $S_i\in I^-$). Therefore, 
\begin{align*}  
\Pr\big(S_i = \win(K),\;S_i\notin Surprise\big)\cdot\mu(S_i)
&\leq \Pr\big(h(S_1)\leq L/\mu(S_i)\big)\cdot
\Pr\big( S_i = \win(I^-)\big)\cdot\mu(S_i)\\
&=
\underbrace{\Big(1 - e^{-L\mu(S_1)/\mu(S_i)}\Big)}_{\leq L\mu(S_1)/\mu(S_i)}
\cdot
\Pr\big( S_i = \win(I^-)\big)\cdot\mu(S_i)\\
&\leq
\Pr\big( S_i = \win(I^-)\big)\cdot L\cdot\mu(S_1).
\end{align*}
We combine all bounds on terms of (\ref{eq:warmup-main}) and get the following bound on the expected cost of the game:
$$\mu(S_1) + 
\sum_{i=2}^k \Pr\big( S_i = \win(I^-)\big)\cdot L\cdot\mu(S_1) + \mu(S_1) = (L+2)\cdot\mu(S_1) = (\ln k+2) \cdot \mu(S_1).
$$
This concludes the proof of the theorem for the case when $S_1$ does not overlap with $S_2,\dots,S_k$. 
We now analyze surprise sets.
\subsection{Surprise Sets}\label{sec:surprise}
In this section, we prove a bound on the probability that a set $S_i$ is a surprise set. We no longer assume that $S_1$ does not intersect with other sets $S_i$. We first show a lemma about exponential random variables.

\begin{lemma}\label{lem:min-expon-rv}
Let $X$ and $Y$ be two independent exponential random variables with positive parameters $\lambda_X$ and $\lambda_Y$, respectively. Then, for every $T\geq 0$, we have
\begin{equation}
\label{eq:min-exp-rv}
\Pr\big(Y \geq X \geq T\big)
= \frac{\lambda_X}{\lambda_X + \lambda_Y}\cdot e^{-(\lambda_X + \lambda_Y)T}.
\end{equation}
\end{lemma}
\begin{proof}
The desired probability can be easily found by  computing $\int_{T}^{\infty} (F_X(t) - F_X(T))f_Y(t) dt$, where 
$F_X(t) = 1-e^{-\lambda_X t}$ is the cumulative distribution function of $X$, and $f_Y(t)=\lambda_Y\cdot e^{-\lambda_Y t}$ is the probability density function of $Y$. Here, we give an alternative proof.  Write,
\begin{align*}
\Pr\big(Y \geq X \geq T\big)
&=
\Pr\big(Y\geq X \text{ \& } 
\min(X,Y)\geq T\big)\\
&=
\Pr\big(X \leq Y\mid \min(X,Y)\geq T)\cdot
\Pr\big(\min(X,Y)\geq T\big).
\end{align*}
We have
$\Pr\big(\min(X,Y) \geq T\big) = e^{-(\lambda_X + \lambda_Y)T}$, because the minimum of two independent exponential random variables with parameters $\lambda_X$ and $\lambda_Y$ is an exponential random variable with parameter 
$\lambda_X+\lambda_Y$. Then,
$\Pr\big(X \leq Y\mid \min(X,Y)\geq T) = \Pr\big(X \leq Y)$ because the exponential distribution is memoryless; and $\Pr\big(X \leq Y) = \lambda_X/(\lambda_X + \lambda_Y)$.
\end{proof}

\begin{lemma}\label{lem:pr-surprise-set}
For every set $S_i$, we have
$$
\Pr(S_i\text{ is surprise set})
\leq \frac{1}{k}\cdot \frac{\mu(S_1)}{\mu(S_i)}.$$
\end{lemma}

\begin{proof}
First, we show that $ \min(e(S_i), h(S_1))\leq h(S_i\setminus S_1)$.
\begin{claim}\label{cl:g-geq-min-e-h}
We always have
$\min(e(S_i), h(S_1))\leq h(S_i\setminus S_1)$.
\end{claim}
\begin{proof}
Consider an arbitrary realization of the game and the time $t=h(S_i\setminus S_1)$ when $S_i\setminus S_1$ is hit. If by this time, $S_1$ has already been hit then $h(S_1)< t$. Similarly, if by this time, $S_i$ has already been eliminated then $e(S_i)< t$. Otherwise, both $S_1$ and $S_i$ are still remaining in the game at time $t$. Therefore, when we pick $\omega \in S_i\setminus S_1$ at time $t$, set $S_i$ gets eliminated (since $\omega\in S_i$; $\omega\notin S_1$; both $S_1$ and $S_i$ are remaining in the game). Thus, in this case, $e(S_i)=t$. This concludes the proof.
\end{proof}

If $S_i$ is a surprise set, then
$\min(e(S_i),h(S_1))=h(S_1)\geq L/\mu(S_i)$. By Claim~\ref{cl:g-geq-min-e-h}, we have
$$
h(S_i\setminus S_1)
\geq \min\big(e(S_i),h(S_1)\big) = h(S_1) \geq L/\mu(S_i).
$$
Thus,
$$
\Pr(S_i\text{ is surprise set}) \leq
\Pr\Big(h(S_i\setminus S_1)\geq h(S_1)\geq L/\mu(S_i)\Big).
$$
By Lemma~\ref{lem:min-expon-rv} applied to the independent exponential random variables 
$h(S_1)$, $h(S_i\setminus S_1)$, and time $T=L/\mu(S_i)$, we have
$$\Pr(S_i\text{ is surprise set})
\leq
\frac{\mu(S_1)}{\mu(S_i\setminus S_1) + \mu(S_1)}\cdot e^{-\frac{L(\mu(S_i\setminus S_1)+\mu(S_1))}{\mu(S_i)}}
\leq \frac{1}{k}\cdot \frac{\mu(S_1)}{\mu(S_i)}.
$$
\end{proof}

\subsection{General Case}\label{sec:general-case}
\begin{proof}[Proof of Theorem~\ref{thm:main}]
We upper bound the expected cost of the game for arbitrary sets $S_1,\dots,S_k$. As before, we assume that $S_1$ is the smallest set. We remind the reader that each hitting time $h(S_i)$ is an exponential random variable with parameter $\mu(S_i)$. In the proof, we will use the definitions of surprise sets (see Definitions~\ref{def:surprise}). We also set $L = \ln k$.

We separately upper bound the cost of the winner depending on whether the winner is 
(a) set $S_1$, 
(b) surprise set,
(c) non-surprise set. Write
\begin{align}
\tag{a}
\E\big[\mu(\win(K))\big] &= 
\E\big[\mu(\win(K))\cdot \ONE\{\win(K) = S_1\}\big]\\
&+ 
\tag{b}
\E\big[\mu(\win(K))\cdot \ONE\{\win \text{ is surprise set}\}\big]\\
&+ 
\tag{c}
\E\big[\mu(\win(K))\cdot \ONE\{\win \text{ is non-surprise set}\}\big].
\end{align}
Term (a) is upper bounded by $\mu(S_1)$. 
We bound term (b) using Lemma~\ref{lem:pr-surprise-set}: The probability that a set is a surprise set is at most $\nicefrac{1}{k}\cdot \mu(S_1)/\mu(S_i)$. Thus, the expected total measure of all sets (not only the surprise winner) is upper bounded by
$\frac{1}{k}
\sum_{i=2}^k \frac{\mu(S_1)}{\mu(S_i)}
\mu(S_i) <
\mu(S_1)$. 

We now bound term (c). Define a new random variable: Let
$\cost(\omega)$ be the cost of the winner (i.e., $\mu(S_i)$, where $S_i$ is the winner) if (1) the winner is a non-surprise set, and (2) $\omega$ is the first element that was chosen in $S_1$. We let $\cost(\omega) = 0$, otherwise. If $\omega$ is the first element that was chosen in $S_1$, then $h(S_1) = h(\omega)$. So, the definition of $\cost(\omega)$ can be written as follows:
$$
\cost(\omega) = \mu(\win(K))\cdot \ONE\{h(S_1) = h(\omega)\}\cdot \ONE\{\win(K) \not\in Surprise\}.
$$
Since the hitting time $h(S_1)$ is finite with probability $1$, the term (c) equals
$$(c) = 
\sum_{\omega\in S_1} \E[\cost(\omega)].
$$
Lemma~\ref{lem:bound-cost-main}, which we prove below, gives a bound of $2L\mu(S_1)$ on the expression above. 
Combining upper bounds on terms (a), (b), and (c), we get 
$$
\E\big[\mu(\win(K))\big]\leq (1 + 2L +1)\mu(S_1) = 
(2\ln k + 2) \cdot \mu(S_1).
$$
\end{proof}

\begin{lemma}\label{lem:bound-cost-main}
For every $\omega\in S_1$, we have
$
\E[\cost(\omega)] \leq 2L\mu(\omega) 
$.
\end{lemma}
\begin{proof}
We have
\begin{equation}
\E[\cost(\omega)] =\E\Big[\mu(\win(K))\cdot \ONE\{h(S_1) = h(\omega)\}\cdot \ONE\{\win(K) \not\in Surprise\}\Big].
\label{main:line:expensive}
\end{equation}
If $S_i$ is a non-surprise set, then $h(S_1) < L/\mu(S_i)$ or $e(S_i) < h(S_1)$. If $S_i$ is the winner, then 
$e(S_i) \geq h(S_1)$. Thus, if $S_i$ is a non-surprise winner, then $h(S_1) < L/\mu(S_i)$. This observations gives us the following upper bound on~(\ref{main:line:expensive}):
\begin{equation}
\label{eq:short-bound-expensive}
\E\big[\cost(\omega)\big] \leq
\sum_{i=2}^k
\mu(S_i)\cdot \Pr\Big(
S_i = \win(K)\text{ and }
h(\omega) = h(S_1) < L/\mu(S_i)\Big).
\end{equation}
Define two set systems $I^-_{\omega}$ and $I^+_{\omega}$ of sets $S_i$ containing and not containing $\omega$:
\begin{align*}
I^-_{\omega} &= 
\{S_i: \omega \notin S_i\text{ and }i\geq 2\};\\
I^+_{\omega} &= 
\{S_i: \omega \in S_i\text{ and }i\geq 2\}.
\end{align*}
Note that $K\equiv \{S_1,\dots,S_k\} = \{S_1\}\cup I^-_{\omega} \cup I^+_{\omega}$. By Lemma~\ref{lem:winner-in-set-winners}, 
$$\win(K) \in 
\big\{S_1,\win(I^-_{\omega}),\win(I^+_{\omega})\big\}.
$$
Observe that if $S_i$ with $i\geq 2$ is the winner, then $S_i=\win(I^-_{\omega})$ or $S_i=\win(I^+_{\omega})$. 
We replace the condition $S_i=\win(K)$ with $S_i\in\{\win(I^-_{\omega}),\win(I^+_{\omega})\}$ in (\ref{eq:short-bound-expensive}) and get bound:
$$
\E\big[\cost(\omega)\big] \leq
\sum_{i=2}^k
\mu(S_i)\cdot \Pr\Big(
S_i \in \{\win(I^-_{\omega}),\win(I^+_{\omega})\}
\text{ and }
h(\omega) < 
\frac{L}{\mu(S_i)}\Big).
$$
The key observation now is that sets $\win(I^-_{\omega})$ and $\win(I^+_{\omega})$ are independent of $h(\omega)$. This is the case, because sets remaining in the competitions $\calR_n(I^-_{\omega})$ and $\calR_n(I^+_{\omega})$ do not change when we select $\omega$. The set 
$\calR_n(I^-_{\omega})$ does not change in the round $n$ when $\omega$ is chosen because all sets $S_i$ in $\calR_n(I^-_{\omega})\subset I^-_{\omega}$ do not contain $\omega$. The set 
$\calR_n(I^+_{\omega})$ does not change in this round because all sets $S_i$ in $\calR_n(I^+_{\omega})\subset I^+_{\omega}$ contain $\omega$ and consequently when $\omega$ is chosen, none of these sets is removed from  $\calR_n(I^+_{\omega})$ (otherwise, $\calR_n(I^+_{\omega})$ would become empty).
Thus,
$$
\E\big[\cost(\omega)\big] \leq\sum_{i=2}^k \mu(S_i)\cdot \Pr\big(S_i \in \{\win(I^-_{\omega}),\win(I^+_{\omega})\}\big)
\cdot 
\Pr\Big(h(\omega) < \frac{L}{\mu(S_i)}\Big).
$$
Using that $h(\omega)$ is an exponential random variable with parameter $\mu(\omega)$, we get (for every $i$)
$$
\mu(S_i)\cdot \Pr\Big(h(\omega)\leq \frac{L}{\mu(S_i)}\Big)
=
\mu(S_i)\cdot \Big(1- e^{-L\frac{\mu(\omega)}{\mu(S_i)}}\Big)
\leq 
\mu(S_i)\cdot L\frac{\mu(\omega)}{\mu(S_i)} = \mu(\omega) L
.$$ 
Hence,  
$$
\E\big[\cost(\omega)\big] \leq
\mu(\omega)L\cdot \sum_{i=2}^k \Pr\big(S_i \in \{\win(I^-_{\omega}),\win(I^+_{\omega})\}\big).
$$
The sum on the right hand side is at most $2$. Thus,
$\E[\cost(\omega)]\leq 2L\mu(\omega)$.
\end{proof}

\bibliographystyle{plainnat}
\bibliography{references}

\end{document}